\documentclass[graphicx]{pasj00}
\Received{$\langle$2009 June 25$\rangle$}
\Accepted{$\langle$2010 February 8$\rangle$}
\Published{$\langle$2010 April 25$\rangle$}
\SetRunningHead{Evidence of Large-Scale Winds}{Tsujimoto et al.}

\usepackage{times}

\def\ltsima{$\; \buildrel < \over \sim\;$}
\def\ltsim{\lower.5ex\hbox{\ltsima}}
\def\gtsima{$\; \buildrel > \over\sim \;$}
\def\gtsim{\lower.5ex\hbox{\gtsima}}
\def\ms{$M_{\odot}$ }
\def\msp{$M_{\odot}$}

\begin{document}

\title{Evidence of Early Enrichment of the Galactic Disk by Large-Scale Winds}
\author{Takuji Tsujimoto, \altaffilmark{1} Joss Bland-Hawthorn, \altaffilmark{2} Kenneth C. Freeman, \altaffilmark{3}}
\altaffiltext{1}{%
   National Astoronomical Observatory of Japan, 2--21--1 Osawa,\\
   Mitaka-shi, Tokyo 181--8588}
\email{taku.tsujimoto@nao.ac.jp}
\altaffiltext{2}{%
   Sydney Institute for Astronomy, School of Physics, University of Sydney, NSW 2006, Australia
   }
\altaffiltext{3}{%
   Research School of Astronomy and Astrophysics (RSAA), Australian National University, \\ 
   Cotter Road, Weston Creek, ACT 2611, Australia
}
\KeyWords{Galaxy: bulge --- Galaxy: disk --- Galaxy: evolution --- ISM: jets and outflows --- stars: abundances}

\maketitle

\begin{abstract}
Large-scale homogeneous surveys of Galactic stars may indicate that the elemental abundance gradient evolves with cosmic time, a phenomenon that was not foreseen in existing models of Galactic chemical evolution (GCE). If the phenomenon is confirmed in future studies, we show that this effect, at least in part, is due to large-scale winds that once enriched the disk. These set up the steep abundance gradient in the inner disk ($R_{\rm GC}$\ltsim $14$ kpc). At the close of the wind phase, chemical enrichment through accretion of metal-poor material from the halo onto the disk gradually reduced the metallicity of the inner region, whereas a slow increase in the metallicity proceeded beyond the solar circle. Our ``wind+infall" model accounts for flattening of the abundance gradient in the inner disk, in good agreement with observations.  Accordingly, we propose that enrichment by large-scale winds is a crucial factor for chemical evolution in the disk.  We anticipate that rapid flattening of the abundance gradient is the hallmarks of disk galaxies with significant central bulges.
\end{abstract}

\section{Introduction}

How do disk galaxies form and how do they evolve through cosmic time? Clear answers to these questions continue to elude us, and it may be many years before we converge on a successful physical model. Disks lie at the forefront of galaxy formation and evolution, not least because most stars are in disks today \citep{Benson_07, Driver_07}. It is widely recognized that N-body simulations of galaxy formation within cold dark matter (CDM) cosmology fail to produce realistic galactic disks \citep{Navarro_94, Steinmetz_95}. But a useful aspect of these incomplete models has been to highlight the possible role of feedback in shaping galaxies. Vigorous feedback in the early phase of galaxy formation is a partial solution to the angular momentum problem by preventing baryons from losing their specific angular momentum too much through the interaction with dark matter \citep{Fall_02}. Indeed, some authors have claimed more realistic disks retaining much more of their angular momentum when one includes the action of AGN-driven jets \citep{Robertson_06} or starburst driven winds \citep{Efstathiou_00}.

There are several new observations that suggest outflows are important in the life cycle of galaxies. First, \citet{Tremonti_04} find evidence for chemical enrichment trends throughout star-forming galaxies over three orders of magnitude in stellar mass. Surprsingly, the enrichment trends are observed to continue all the way down to $log(L/L_\odot$) = 4 \citep{Kirby_08}. Indeed, there is tentative evidence that galaxies that exceed the mass of the Galaxy manage to retain a large fraction of their metals, in contrast to lower mass galaxies where metal loss appears to be anticorrelated with baryonic mass. Recent theoretical work has shown that a low-metal content in dwarf galaxies is attributable to efficient metal-enriched outflows \citep{Dalcanton_07}, although this has other possible interpretations \citep{Tassis_08}. The intergalactic medium (IGM) at all redshifts shows signs of significant metal enrichment consistent with the action of winds \citep{Cen_99, Madau_01, Ryan_06, Dave_08}.  A high proportion of Lyman break galaxies at $z\sim 3-4$ show kinematic signatures of winds \citep{Erb_06}. These large-scale outflows are thought to be powered by the activity in the central regions (e.g. bulges) of galaxies. Since these winds entrain a great quantity of heavy elements, there should be signatures of large-scale winds on the disk stellar population.

There is now strong evidence for large-scale outflows in the Galaxy across the electromagnetic spectrum \citep{Bland_03, Fox_05, Keeney_06, 
Everett_10}.
Furthermore, there is tantalizing direct evidence that dust and gas from the inner Galaxy have been transported to the solar neighbourhood \citep{Clayton_97}.
In standard chemical evolution models, the isotopes $^{29}$Si and $^{30}$Si become more abundant than $^{28}$Si as the disk ages. But pre-solar
grains from meteorites show evidence that the local isotopes formed in material that had experienced more nuclear synthesis than material that came 
after the Sun's birth. 

\begin{figure*}
\vspace{0.2cm}
\begin{center}
\FigureFile( 107mm, 107mm ) {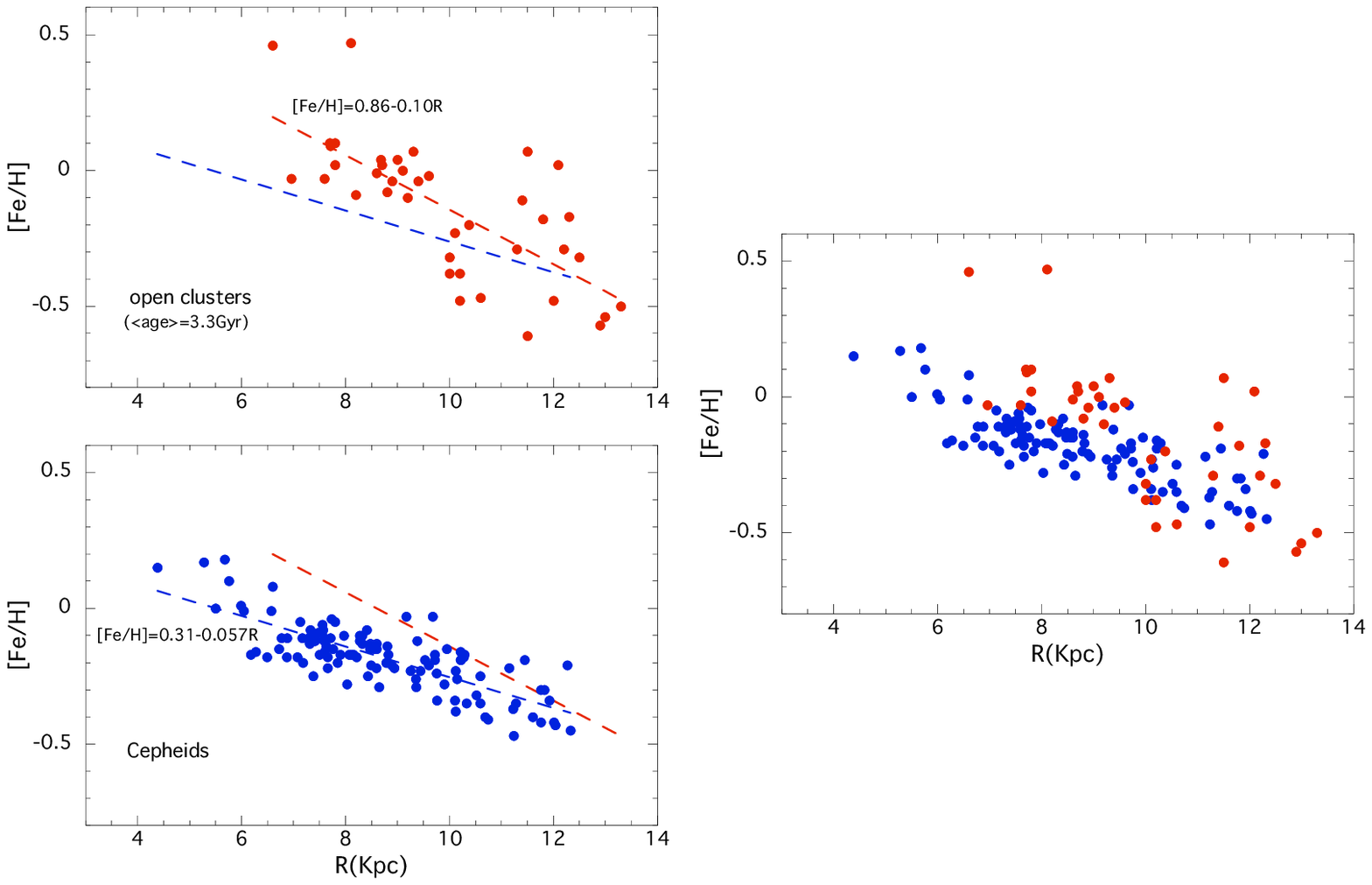}
\end{center}
\vspace{0.3cm}
\caption{{\it left panels}:  The observed Fe abundances as a function of Galactocentric distance $R$ within 14 kpc for open clusters with their ages older than 1 Gyr, tabulated in Table 1 (upper), and those for Cepheids (q.v., \cite{Andrievsky_04}) (lower). The [Fe/H] values for Cepheids are shifted by -0.16 dex as performed by \citet{Yong_06}. In each panel, the line of the least-squre fitting for the other  population is shown for reference. {\it right panel}: The comparison of two distributions shown in left panels.}
\end{figure*}

This body of evidence makes it crucial to consider the influence of large-scale winds within Galactic chemical evolution (GCE) models. The key observational constraint here is the Galactic abundance gradient determined from different astrophysical sources (e.g., K giants, Cepheids). Simple GCE models in which the disk is divided into independently evolving rings successfully reproduce the present abundance gradient in the disk (e.g., \cite{Boissier_99, Hou_00, Chiappini_01}). The same holds true for models that consider the redistribution of elements in the disk through radial flows (e.g., \cite{Lacey_85, Tsujimoto_95, Portinari_00}).

None of the published GCE models to date predicted the time evolution of abundance gradient 
recently established in large homogeneous stellar surveys. The observed gradient has flattened out by half in the last 5 Gyr or so (e.g., \cite{Chen_03, Maciel_06}). Such a large change in the abundance gradient seems beyond the capabilities of GCE models to date such that a new mechanism must be considered, as we show. In addition, the chemodynamical model which first considers the influence of winds from the bulge on the disk during the collapsing stage of a galaxy \citep{Samland_97} predicts a gradual steepening of abundance gradient over the last 9 Gyr.

\section{Observed Features on Time-Dependent Abundance Gradient}

In this section, we review evidence for the time evolution of the Galactic abundance gradient. The best studied elements for the discussion of abundance gradient are O and Fe, and accordingly their present gradients are determined by the observations of young objects such as H II regions or OB-type stars for O (e.g., \cite{Esteban_05, Rolleston_00, Daflon_04}) and Cepheids for Fe (e.g., \cite{Andrievsky_04, Luck_06}). However, the crucial issue is the determination of their past gradients. The corresponding old objects for each element are planetary nebulae (PNe) (e.g., \cite{Maciel_06, Perinotto_06}) and open clusters (OCs) \citep{Friel_02, Chen_03, Bragaglia_06}, respectively. 

To date, there is a very wide literature on both O and Fe in stars. But we confront the fact that age-dating of old stars makes it difficult to deduce the time evolution of the abundance gradient when compared to younger objects, because the ages of both PNe and OCs are widely distributed from $10^6$ yr to of $10^9$ yr. In this regard, OCs whose ages are inferred from their color-magnitude diagrams are much more suitable targets than PNs for this study, though some effort to deduce the ages of PNs has been done \citep{Maciel_03}. Thus, we utilize the Fe abundance of Cepheids and OCs to discuss the time evolution of abundance gradient.

Another reason for focusing on the [Fe/H] gradient, rather than [O/H], is attributable to our theoretical framework that considers the effect of large-scale outflows on the chemical evolution of the disk. As discussed in the next section, it is likely that winds from the bulge are not so metal-rich in O. Thus, O is not a good indicator for the wind enrichment that is predicted to cause the time evolution of the abundance gradient in our scheme.  Therefore, we choose Fe to study the time evolution of the abundance gradient because of the improved stellar ages, and the ease of implementation in our models.

We have compiled data for OCs with ages older than 1 Gyr, and their age, Galactocentric distance $R$, and [Fe/H] are tabulated in Table 1 (Appendix). The mean age of these OCs is 3.3 Gyr. First, we plot  the [Fe/H] abundances as a function of $R$ for the OCs within 14 kpc in the left upper panel of Figure 1. The least-square fitting gives a slope of $d$[Fe/H]/$dR$=-0.10 dex kpc$^{-1}$. There exist two very metal-rich OCs at the inner part, i.e., NGC 6253 with [Fe/H]=+0.46 \citep{Carretta_07} and NGC 6791 with [Fe/H]=+0.47 \citep{Gratton_06}, which seemingly make the gradient steeper. However, these OCs are not the determining factor to fix the slope. In fact, fitting the data with or without these OCs gives a similar slope of -0.09 dex kpc$^{-1}$. On the other hand,  the [Fe/H] gradient for Cepheids (q.v., \cite{Andrievsky_04}), the majority of which have  the ages of an order of $10^7$ yr, is shown in the left lower panel. The shallower slope of -0.057 dex kpc$^{-1}$ is deduced, and this value is in good accord with that (-0.06 dex kpc$^{-1}$) derived by \citet{Luck_06} and -0.052 dex kpc$^{-1}$ by \citet{Lemasle_09}. In addition, if we choose Mg as an alternative element [recalling that O should be avoided], OB stars give the present-day gradient of -0.052 dex kpc$^{-1}$ \citep{Daflon_04}. Results of the least-square fittings with one sigma errors are as follows.

\vspace{0.3cm}
{\sl open clusters ($<$age$>$=3.3 Gyr)}

\begin{eqnarray}
{\rm [Fe/H]}=0.88 (\pm 0.15) -0.10 (\pm 0.015) R \nonumber
\end{eqnarray}

\vspace{0.3cm}
{\sl Cepheids}

\begin{eqnarray}
{\rm [Fe/H]}=0.31 (\pm 0.037) -0.057 (\pm 0.004) R \nonumber
\end{eqnarray}
\vspace{0.1cm}

The comparison of the two measurements confirms the flattening of the [Fe/H] gradient, and its rate is estimated to be 0.013 dex kpc$^{-1}$ Gyr$^{-1}$. Note that the rates obtained by the previous studies are 0.0094 dex kpc$^{-1}$ Gyr$^{-1}$ for the [O/H] abundance of PNe \citep{Maciel_07}, and 0.008 dex kpc$^{-1}$ Gyr$^{-1}$ which is deduced from the comparison of young and old OCs (\cite{Chen_03}; see also Magrini et al. 2009). As mentioned in the previous section, such a rate of flattening cannot arise from GCE models. Our result suggests the observed change in [Fe/H] gradient during the last 5 Gyr should be about  0.065 dex kpc$^{-1}$, whereas the corresponding predictions are in the range 0.02 dex kpc$^{-1}$ \citep{Boissier_99, Hou_00} and -0.01 dex kpc$^{-1}$ (minus means a steepening; \cite{Chiappini_01}). Moreover, the right panel showing the comparison of two distributions seems to indicate that a flattening of abundance gradient in the last few Gyr is mainly caused by a decrease in [Fe/H] of the inner disk.

\section{"Wind + Infall" model}

We try to reproduce a flattening of abundance gradient across the disk by incorporating the enrichment by large-scale winds into the model. The basic picture is that the disk was formed through a continuous low-metal infall of material from outside the disk region based on the inside-out formation scenario \citep{Matteucci_89}, that is, the disk is formed by an infall of gas occurring at a faster rate in the inner region than in the outer ones
consistent with the shorter dynamical times. For a specific period, heavy elements carried by winds from the bulge dropped and enriched the disk. Here we calculate chemical evolution at three regions with their Galactocentric distances $R$=4, 8 (solar vicinity), and 12 kpc.

Let $\psi(t)$ be the star formation rate and $A(t)$, $w(t)$ be the gas infall rate, the wind rate, respectively,  then the gas fraction $f_g(t)$ and the abundance of heavy-element $i$ $Z_i(t)$ in the gas at each region change with time according to 
\begin{equation}
\frac{df_g}{dt}=-\alpha\psi(t)+A(t)+w(t)
\end{equation}
\begin{eqnarray}
\frac{d(Z_if_g)}{dt}=-\alpha Z_i(t)\psi(t)+Z_{A,i}(t)A(t)+y_{{\rm II},i}\psi(t) \nonumber \\
 +y_{{\rm Ia},i}\int^t_0 \psi(t-t_{\rm Ia})g(t_{\rm Ia})dt_{\rm Ia}+W_i(t) \ \ ,
\end{eqnarray}
\noindent where $\alpha$ is the mass fraction locked up in dead stellar remnant and long-lived stars, $y_i$ is the heavy-element yield from an SNII or SN Ia, and $Z_{A,i}$ is the abundance of heavy element  contained in the infalling gas. 

\begin{figure}
\vspace{0.2cm}
\begin{center}
\FigureFile( 60mm, 60mm ) {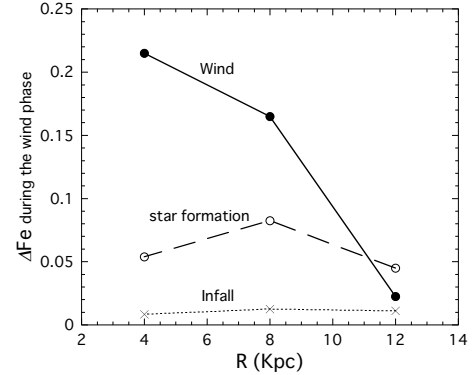}
\end{center}
\vspace{0.3cm}
\caption{The increase in Fe during the wind phase by the wind, the local star formation, and an infall. Each corresponds to the integration of a fifth term (i.e., $W_i(t)$), third and fourth terms, and a second term in the right hand of Eq.[2] over $T_G=$4-6 Gyr, respectively. All values are normalized by the solar Fe abundance. }
\end{figure}

\begin{figure*}
\vspace{0.2cm}
\begin{center}
\FigureFile( 130mm, 130mm ) {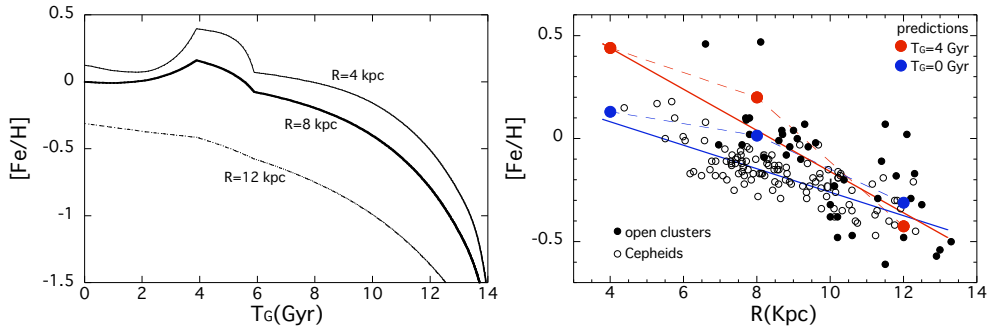}
\end{center}
\vspace{0.3cm}
\caption{{\it left panel}: Predicted age-[Fe/H] relations at the Galactocentric distances of $R$=4, 8, and 12 kpc. The age of galaxy is assumed to be 14 Gyr. See the text for the details. {\it right panel}: Observed [Fe/H] distributions as a function of the distance from the center for Cepheids (open circles; \cite{Andrievsky_04}) and open clusters (filled circles; Table 1), compared with the predicted values at $T_G=4$ Gyr (red circles) and the present (blue circles), which are extracted from the results of the left panel. The observationally implied [Fe/H] gradients from a least square fitting of data are assigned by solid lines for the present (blue: -0.057 dex kpc$^{-1}$) and several Gyr ago (red: -0.1 dex kpc$^{-1}$). 
}
\end{figure*}

The star formation rate $\psi(t)$ is assumed to be proportional to the gas fraction with a constant coefficient of $1/\tau_{\rm SF}(R)$,  where $\tau_{\rm SF}(R)$ is a timescale of star formation as a function of $R$. For the infall rate, we adopt the formula that is proportional to $t\exp(-t/t_{\rm in}(R))$ with a timescale of infall $\tau_{\rm in}(R)$. According to the inside-out scenario, $\tau_{\rm in}(R)$ and  $\tau_{\rm SF}(R)$ are assumed to increase outwards, and the adopted timescales in unit of Gyr are ($\tau_{\rm in}$, $\tau_{\rm SF}$) =(2.5, 0.5), (4, 1.2), and (7, 7), respectively. The metallicity $Z_{A,i}$ of an infall is assumed to be time-dependent, which is implied by the cosmic evolution of damped Ly$\alpha$ systems \citep{Wolfe_05}, then $Z_{A,i}(t)$ is given by interpolating four reference points, that is,  [Fe/H] = -2 at $T_G$=14 Gyr, [Fe/H]=-1.5 at $T_G$=10 Gyr, and [Fe/H]=-1 at $T_G$=6 Gyr and the present, with a SN-II like enhanced [$\alpha$/Fe] ratio for all duration. For the initial mass function (IMF), we assume a power-law mass spectrum with a slope of -1.35, which is combined with the nucleosynthesis yields of SNe II and Ia taken from \citet{Kobayashi_06} and \citet{Iwamoto_99}, respectively to deduce $y_{{\rm II},i}$ and $y_{{\rm Ia},i}$. We apply an instantaneous recycling approximation to SN II progenitors, leading to the ejection rate of heavy elements, which is proportional to $\psi(t)$. In contrast, the effect of time delay yields the ejection rate from SN Ia's, which is proportional to $\psi(t-t_{\rm Ia})$. It is assumed that $t_{\rm Ia}$ spans over some range according to a distribution function $g(t_{\rm Ia})$.  Here we assume that the fraction of the stars that eventually produce SNe Ia for $3-8$\ms is 0.05 with a box-shaped $g(t_{\rm Ia})$ for $0.5\leq t_{ Ia}\leq3$ Gyr \citep{Yoshii_96}. Recent studies on the SN Ia rate in distant galaxies imply the different form of $g(t_{\rm Ia})$ favoring a large population of young progenitors for SNe Ia \citep{Mannucci_06, Sullivan_06, Totani_08}. However, such a SN Ia frequency distribution is evidently at odds with the presence of an observed plateau of [$\alpha$-element/Fe] among halo stars. Thus, here we adopt the above form so that our model can reproduce the elemental features of halo stars.

$W_i(t)$ is an accretion rate of heavy elements entrained in winds. Heavy elements in winds are added to each region of the disk in a sense that they drop more in a place closer to the center \citep{Bekki_09}. We assume a constant rate $W_i(R)$ with time as a function of $R$. Since the winds triggered by starbursts are likely not to entrain the ISM in the bulge so much, as implied by an enhanced [$\alpha$/Fe] ratio in dwarf starburst winds \citep{Martin_02}, we determine the vales of $w(R)$ so that the Fe abundance in winds is adjusted to [Fe/H]=$+0.4$, which is identical to the metallicity of super metal-rich stars in the solar neighborhood. Accordingly, relative ratios among elements in winds are assumed to be equivalent to those in a SN II ejecta for the Mg/Fe ratio. On the other hand, we assume a low O abundance in the wind since this is implied by the bulge abundance \citep{Fulbright_07}. Here the reduced O yield of SN II is probably due to efficient mass loss for massive stars, which prevents He and C from synthesizing into O within the metal-rich environment in the bulge \citep{McWilliam_08}. Thus, we assume that the O yield of SN II in the wind decreases by a factor of 0.25. The period $P_{\rm wind}$ of the occurrence of winds is set to be $P_{\rm wind}$=$4-6$ Gyr so as to reproduce the chemical features of solar neighborhood stars, and $W_i(R)$=0 ($w(R)$=0) outside this period. This setting is also favorable to the observed flattening in the last several Gyr. The expression of $W_i(R)$ in comparison with the production rate $y_{{\rm II},i}\psi(t)$ of SN II elements by star formation is useful to know the degree of enrichment by winds. Here, we define an enhancement parameter $p(R)$, defined as $p(R)=W_i(R)/(y_{{\rm II},i}<\psi(t)>)$, where $<\psi(t)>$ denotes an average star formation rate at each region during $P_{\rm wind}$. Thus, an increase rate of elements by both star formation and winds is approximated by $y_{{\rm II},i}\psi(t)(1+p(R))$. For the values of $p(R)$, we set $p(R)$=4, 2, and 0.5, respectively. These adopted values of $W_i(R)$ (i.e., $p(R)$) are chosen arbitrarily, adjusted to obtain the satisfactory outcomes. Validation of them must be done with the aid of N-body simulations which will  reveal an accretion rate of the winds as functions of Galactocentric distance.

First, we show in Fig.~2 how the enrichment during the wind phase proceeds through the wind, star formation in situ, and an infall. Little dependence of local enrichment on the distance is owing to the model result that for the inner place a high star formation rate is assumed but a less amount of gas as a result of an efficient gas consumption until the occurrence of wind event, while the opposite situation is realized in the outer part. Accordingly, the steep decrease in enrichment by the wind and thereby a switch of the major contribution from the wind to the local star formation according to the Galactocentric distance lead to a temporally steep abundance gradient which will be followed by its flattening. 

The left panel of Fig.~3 shows the predicted age-metallicity relations at four locations;  on the right, we show the [Fe/H] trends with radius at $T_{\rm G}$=0 (present) and a look-back time of 4 Gyr. Our scenario can be summarized as follows: (i) early wind enrichment sets up the initially steep gradient $-$ this is consistent with IGM enrichment seen to the highest redshifts (Madau et al 2001); (ii) the gradient flattens quite dramatically with time due to large-scale winds from the central starburst; (iii) the IGM gas rains down on the disk from time to time.

Figure 4 shows the resultant abundance distribution function (ADF) and correlations of [Mg/Fe] and [O/Fe] with [Fe/H], which are in good agreement with the observations. It should be stressed that the upturning feature of [Mg/Fe]  for [Fe/H] \gtsim 0 strongly implies the enrichment by large-scale winds having an enhanced Mg/Fe ratio. In addition to it, the subsequent return to the solar abundance as a result of a gradual dilution of the effect of wind enrichment well explains an offset of abundances between the present gas and metal-rich stars, which we discuss in \S 5.1. On the other hand, the downward feature of [O/Fe] is reproduced by the assumed low O/Fe ratio in the wind material.

We now discuss our model in the context of the stellar record in \S 5. 

\begin{figure*}
\vspace{0.2cm}
\begin{center}
\FigureFile( 110mm, 110mm ) {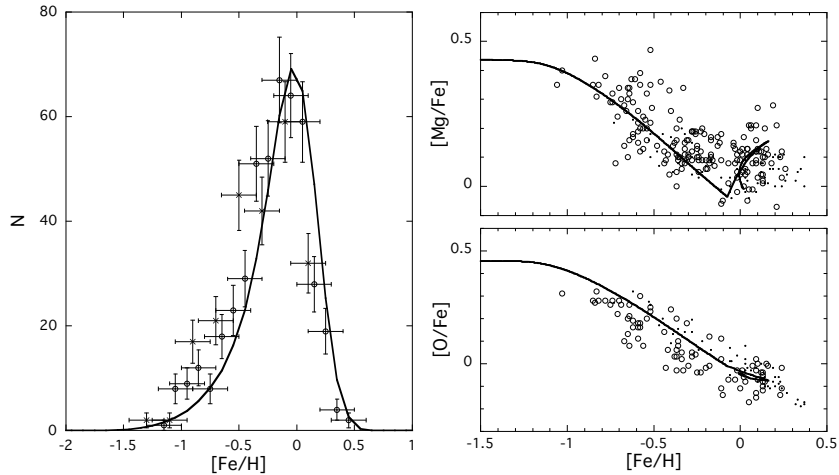}
\end{center}
\vspace{0.3cm}
\caption{Features of chemical evolution in the solar neighborhood. {\it left panel}: Predicted abundance distribution function of thin disk stars against the Fe abundance. The calculated distribution is convolved using the Gaussian with a dispersion of 0.1 dex in [Fe/H]. Open circles and crosses represent data taken from \citet{Edvardsson_93} and \citet{Wyse_95}, respectively.  The model distribution and the observed one by \citet{Wyse_95} are normalized to coincide with the total number of the sample stars used by \citet{Edvardsson_93}.{\it right panel}: Correlation of [Mg/Fe] and [O/Fe] with [Fe/H] for thin disk stars with our ``wind+infall" model superimposed. The open circles and dots are taken from \citet{Edvardsson_93} and \citet{Bensby_05}, respectively.
}
\end{figure*}

\section{Link with Chemical Evolution of the Bulge}

\begin{figure}
\vspace{0.2cm}
\begin{center}
\FigureFile( 50mm, 80mm ) {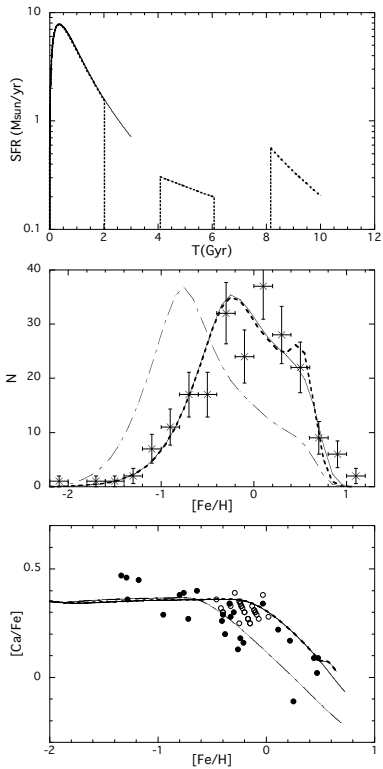}
\end{center}
\vspace{0.3cm}
\caption{Features of chemical evolution in the Galactic bulge. {\it upper panel}: Star formation history for the model with one initial starburst (model 1; solid line) and with three episodes of star formation (model 2; dotted line). {\it middle panel}: Abundance distribution function of bulge stars against the Fe abundance. Solid and dotted lines are the predictions of the model 1 and 2 with a  flatter IMF ($x$=1.05). Result of the model 1 with a Salpteter IMF ($x$=1.35) is shown by dash-dotted line. These calculated distributions are convolved using the Gaussian with a dispersion of 0.1 dex in [Fe/H]. Crosses with error bars represent data taken from \citet{Fulbright_06}. {\it lower panel}: Correlation of [Ca/Fe] with [Fe/H] for bulge stars. Three lines are the same as for middle panel. These model results are compared with the observational data (filled circles: Fulbright et al. 2007; open circles: Rich \& Origlia 2005, Rich et al. 2007).
}
\end{figure}

A driver of large-scale winds would be a starburst in the Galactic bulge. Therefore, our scenario proposes that the bulge has a complex history involving at least one major starburst episode around 5 Gyr ago in addition to the initial burst of star formation. To date, the star formation history of the Galactic bulge still remains an open question because we do not have the precise color-magnitude diagram (CMD) of the bulge owing to the difficulty of discrimination of bulge stars from disk stars (e.g., \cite{Holtzman_93, Feltzing_00}). Here we discuss whether the scheme based on the episodic star formation as proposed here is compatible with the observed chemical properties of bulge stars. Two cases are considered: one is that all bulge stars are formed in an initial starburst (model 1), and the other is that the bulge experiences three episodes of star formation (model 2). To see the case where a star formation exclusively influences the chemical evolution of the bulge, all the metal from SNe are assumed to be consumed for enrichment process inside the bulge. The common assumption for both models is that the period of star formation is set to be in an order of Gyr since the elemental features of bulge stars \citep{Fulbright_07} show a clear signature of SNe Ia (see the lower panel of Fig.~5). Accordingly, we assume that a period of star formation is 3 Gyr in model 1, and 2 Gyr for individual episodes that produce 88\%, 5\%, and 7\% of stars, respectively, in model 2. The star formation histories for both models are shown in the  upper panel of Figure 5. 

First, we show that the model with an enhanced star formation is not sufficient to reproduce the chemical feature of bulge stars. The dash-dotted line in the middle panel represents a resultant ADF calculated by the model 1 with a rapid collapse ($\tau_{\rm in}$=0.2 Gyr) and a high SFR ($\tau_{\rm SF}$=0.5 Gyr) together with a Salpeter IMF. The ADF thus obtained is entirely skewed to a low-metallicity. In addition, the predicted [Ca/Fe] curve in the lower panel is lower than the observed data in a metal-rich regime (Ca is chosen among $\alpha$-elements instead of O or Mg that are conventionally used since the bulge stars exhibit an incompatible elemental features between [O/Fe] and [Mg/Fe]; see Fulbright et al.~2007).  These inconsistencies are resolved by the models with a flatter IMF ($x$=1.05), as shown by solid (model 1) and dotted lines (model 2), both of which give a good agreement with the observed ADF as well as the observed correlation of [Ca/Fe] with [Fe/H]. This flat IMF is fairly consistent with those claimed by several authors, such as $x=1.1$ \citep{Matteucci_90} and $x=0.95$ (Ballero et al.~2007a,b), and yields a high rate of SNe II, suggesting that the Galactic bulge possesses a potential environment that the powerful winds triggered by numerous SNe II are inclined to take place.

It is difficult to conjecture how massive a starburst is likely to be from a viewpoint of a wind enrichment on the disk since there are too many uncertain factors to predict it. Here we take a simplified case and show that a starburst which produce $\sim 5\%$ of the bulge stars might be one candidate, by checking supply and demand of Fe. For simplicity, the disk is here considered to be one zone represented by the solar circle. Let us take the masses  of a bulge and a disk and the gas fraction of a disk as $2\times 10^{10}$\msp, $4\times 10^{10}$\msp, and 0.1. The IMF with a slope of $x$= -1.05 together with the star formation producing the fraction of $5\%$ of bulge stars yields $\sim 3\times 10^7$ SNe II that eject $\sim 4\times 10^6$ \ms of Fe, some fraction of which will be, though, retained within the bulge.  On the other hand, the Fe abundance of gas in the disk should increase from [Fe/H]=0 to [Fe/H]=+0.3 by the wind enrichment plus the local enrichment by star fromation. It means that the winds should entrain roughly the same amount of Fe as that already present in the ISM to double the Fe abundance if we neglect the local Fe production.  In the end, the total Fe mass necessary for the additional enrichment on the disk amounts to $4\times 10^{10}\times 0.1 \times 10^{-3} ({\rm solar \ Fe \ abundance}) \sim 4\times 10^6$\msp, equivalent to the former estimate. If we assume that the wind enrichment is restricted to the inner disk including the solar vicinity together with a partial contribution to the winds from SNe Ia and hypernovae that eject high Fe masses, the necessary fraction of stars produced in a burst will significantly decrease like  $<$ 1\%. 

\section{Chemical Signatures of Large-Scale Winds}

\subsection{Solar neighborhood}

The ADF of solar neighborhood disk stars has been studied by many authors (e.g., \cite{Wyse_95,Rocha-Pinto_96,Nordstrom_04}), and the location of its peak at [Fe/H]$_{\rm peak}$$\approx$ -0.15 has been firmly established. On the other hand, the metal-rich end of the ADF extends at least to [Fe/H]$\sim$ +0.2, and the fraction of stars with [Fe/H]$>$0 is roughly 20\%. Spectroscopic observations of elemental abundances for metal-rich disk stars \citep{Feltzing_98,Bensby_05} have confirmed that chemical enrichment in the solar neighborhood continued unabated until [Fe/H]$\sim$ +0.4. \citet{Tsujimoto_07} has claimed that the simultaneous reproduction of both the presence of stars with [Fe/H]\gtsim +0.2 and [Fe/H]$_{\rm peak} <$0 is hard to realize through the conventional scheme of local enrichment via low-metallicity gaseous infall from the halo, and that the presence of supersolar stars is crucial evidence for enrichment by outflows from the bulge. 

There is little evidence for substantial amounts of supersolar gas in the solar neighbourhood. Samples of nearby young stars such as Cepheids \citep{Lemasle_07} or OB stars \citep{Daflon_04} have a mean of around solar, including the stars in the Orion association \citep{Cunha_98}, with little or no offset with the HII region population (e.g., \cite{Simon-Diaz_06}). Therefore, there exists a clear discrepancy between the presence of local metal-rich F/G stars with [Fe/H] up to $\sim$ +0.2-0.4 and the mean metallicity of gas in the local disk as long as we take a view of the standard chemical evolution in which the metallicity increases with time. But this offset can be explained quite naturally in a scheme with early-phase rapid enrichment and subsequent dilution. 
 
In the observed age-metallicity relation of local disk stars \citep{Bensby_05, Reid_07}, there is a dearth of young ages for metal-rich stars, implying that these stars are not formed through recent star formation but probably formed a few Gyr ago. This observed fact is consistent with the offset cited above.
 
\subsection{Time evolution of the abundance gradient}

The radial metallicity gradient roughly from $R_{\rm GC}$=4 to $R_{\rm GC}$=14 kpc has flattened out in the last several Gyr with a change in a slope of $\sim $-0.1 dex kpc$^{-1}$ to $\sim$ -0.05 dex kpc$^{-1}$ as discussed in section 2 (see also \cite{Chen_03, Maciel_06}).  Chemical evolution models predict contradictory time evolution of the metallicity gradient, i.e., a steepening \citep{Chiappini_01} or a flattening \citep{Boissier_99, Hou_00}.  Putting aside these contradictions, existing GCE models predict a monotonic increase in abundances for each region, and the predicted gradient change is small compared with the observations. In Fig. 1, we compare the steep relic gradient observed in the open clusters with that observed in the Cepheids. We note that the gradient became shallower owing to a decrease in metallicity in the inner regions, with a corresponding increase in the outer region. We addressed this trend in our model in \S 3.

\section{Is Radial mixing an alternative explanation?}

Radial mixing of disk stars due to resonant scattering with transient spiral waves has a significant influence on the distribution of abundances over the disk (Sellwood \& Binney 2002; Ro\v{s}kar et al.~2008a,b; S\'{a}nchez-Bl\'{a}zquez et al.~2009). In fact, it provides an excellent explanation for the observed age-metallicity relation in the solar neighborhood whose feature is characterized by an almost flat relation with a large spread \citep{Sellwood_02, Roskar_08b}. It is worthwhile to discuss whether this mechanism can explain the chemical features of the disk raised in this paper. 

\noindent {\sl 1. time evolution of the abundance gradient}

Radial mixing and our wind scenario make very different predictions about how the abundance
gradient evolves with cosmic time. Radial mixing results in a distinct {\it steepening} of the abundance gradient with time (i.e. the opposite of what is observed) owing to the fact that radial migration is more evident in older populations \citep{Roskar_08b, Sanchez_09}. 
As discussed in \S 2, the prediction of a steepening seems at odds with the present Galactic data. 

\noindent {\sl 2. metal-rich stars in the solar neighborhood}

Radial mixing predicts that contaminants coming from the inner disk populate the solar neighborhood, and allows the presence of metal-rich stars beyond the upper limit of a local chemical enrichment \citep{Roskar_08b}. However, since such mixing is inactive over the inner 4 kpc or so, super metal-rich stars whose metallicities range over +0.2 \ltsim [Fe/H] \ltsim +0.4 are not locally expected. Even if we ignore the limit of an effective range for this mechanism, elemental abundances of nearby metal-rich stars cast doubt on this possibility. As already discussed in \S 3, nearby metal-rich stars are represented by an upturn in [$\alpha$/Fe] ratios. On the other hand, stars orbiting near the Galactic center should exhibit a decreasing [$\alpha$/Fe] trend with increasing [Fe/H], as observed in the bulge \citep{Fulbright_07,Melendez_08}. However, this argument needs caution since some authors claimed no upturning feature in nearby metal-rich stars \citep{Chen_08}. 
 
In summary, we believe that the most critical test to judge the rightness of either our scenario or radial mixing will be validation of the time-dependency of the abundance gradient by a more reliable database. Improved data bases will soon become available from infrared surveys that target the disk and the bulge (e.g. APOGEE; \cite{ Allende_08}). 

\section{Conclusions}

The Galactic bulge has a complex history involving at least one starburst episode around 5 Gyr ago in addition to the initial burst that gave rise to the old population $>10$ Gyr ago. There is certainly good evidence for recent star formation in the nuclear region \citep{Launhardt_02}. In addition, from the perspective of bar formation, one might expect some star formation activity in the region of what is now the bulge, about 3 Gyr after the inner disk started to form - this would be at the time when the bar buckling occurs \citep{Athanassoula_08} - it would really stir up any remaining gas in the inner disk. While numerous studies \citep{Ortolani_95, Feltzing_00, Kuijken_02, Zoccali_03, Clarkson_08} agree that the bulk of the bulge formed $>$10 Gyr, a minority population of stars as young as 5 Gyr cannot be ruled out, and a minor burst of star formation as recent as 5 Gyr may have taken place in the bulge. The fraction of stars produced by such a minor burst must be small like $\sim 5$\% as implied by our model so as not to contradict the present CMDs, but in practice, how much of stars  the starburst was likely to produce remains quite unclear. As well, the age of a burst claimed here is not definitive due to a lack of determinant factor of its age-dating. The results from microlensed dwarf and subgiant stars in the Galactic bulge showing the average age of $\sim$ 7 Gyr for metal-rich ([Fe/H]$>$ 0) bulge stars \citep{Bensby_10} may be in favor of our hypothesis. Moreover, there is now good evidence for bursting star formation in the Galaxy's past over most of cosmic time. \citet{Rocha-Pinto_00} find that the disk has experienced a few bursts in the past, using a chromospheric age distribution of dwarf stars. We have examined the impact of this burst history on the large-scale disk.

We confirm the earlier claims that the [Fe/H] abundance gradient has flattened out during the last several Gyr, from the comparison of observed data between Cepheids and open clusters with their ages older than 1 Gyr, and claim that large-scale winds from the Galactic bulge are the crucial factor for its mechanism. Our proposed scenario is (i) winds once set up a steep abundance gradient ($\sim$ -0.1 dex kpc$^{-1}$) owing to the enrichment by heavy elements entrained in the winds, and (ii) later evolution leads to a flattening of abundance gradient through chemical evolution under an accretion of a low-metal gas from the halo. Accordingly, we predict that a flattening is the hallmark of disk galaxies with significant central bulges, and thereby the variation in wind signature among galaxies imprinted on the disk may be correlated with the size of  their bulges.

We do not doubt that the detailed processes involved in Galaxy evolution are highly complex, involving accretion, feedback, stellar evolution and so forth. In addition, recent numerical simulations highlight the possibility of strong secular evolution (e.g. in-plane disk scattering). Here we highlight the possible role of large-scale winds on the stellar record. These need to be factored into numerical simulations of disk formation and evolution, either in the form of a superwind (e.g., \cite{Heckman_90}) or a more quiescent fountain flow through the halo \citep{Efstathiou_00}.

A huge database of stellar elemental abundances is required to unravel the various interlocking issues. We anticipate major progress to come from the new generation of million-star surveys, including HERMES, SEGUE, LAMOST and GAIA, set to launch in the next decade. A key issue is whether and how the chemical evolution of the Galactic disk inside the solar circle during the last several Gyr has deviated from the standard picture predicted by the GCE models. The interrelation of star formation history between the bulge and the disk can be investigated in future surveys, particularly if we obtain accurate abundance measurements for several independent line element groups (e.g. $\alpha$/Fe, CNO). 

Finally, we accept that our paper was missing a key prediction. Our "wind+infall" model anticipates a present-day high D/H abundance in the local ISM, recently revealed by \citet{Linsky_06}, owing to our essential idea that all heavy elements are not locally produced but some fraction comes form the outside. Details will be presented in our forthcoming paper.

\bigskip
JBH is supported by a Federation Fellowship from the Australian Research Council (ARC). This work is assisted in part by Grant-in-Aid for Scientific Research (21540246) of the Japanese Ministry of Education, Culture, Sports, Science, and Technology and by ARC grant DP0988751 that provides partial support for the HERMES project. 

\appendix
\section*{}
\begin{table}[h]
\begin{center}
\caption{[Fe/H], distance, and age of open clusters with age$>$ 1 Gyr}
{\scriptsize 
\begin{tabular}{lccccc}
\hline
\multicolumn{1}{c}{cluster} & \multicolumn{1}{c}{R(kpc)} & \multicolumn{1}{c}{Age(Gyr)} & \multicolumn{1}{c}{[Fe/H]} & \multicolumn{1}{c}{$\sigma$[Fe/H]} & \multicolumn{1}{c}{References} \\
\hline
Be 12 & 11.5 & 4 & 0.07 & - & 1 \\
Be 17 & 11.4 & 10 & -0.11 & 0.12 & 2 \\
Be 18 & 12.09\footnotemark[$*$] & 4.26 & 0.02 & 0.15 & 1 \\
Be 19 & 13.3 & 3.09 & -0.5 & - & 1 \\
Be 20 & 16 & 5 & -0.49 & 0.21 & 3 \\
Be 21 & 12.99 & 2.2 & -0.54 & 0.2 & 4 \\
Be 22 & 14.02 & 2.4 & -0.32 & 0.19 & 4 \\
Be 25 & 18.2 & 5 & -0.20 & 0.05 & 5\\
Be 29 & 20.81 & 3.7 & -0.44 & 0.18 & 4 \\
Be 31 & 12.9 & 4 & -0.57 & 0.23 & 3 \\
Be 32 & 11.3 & 6.5 & -0.29 & 0.04 & 6 \\
Be 39 & 12.3 & 7.94 & -0.17 & 0.09 & 1 \\
Be 64 & 11.5 & 1 & -0.61 & - & 1 \\
Be 66 & 12 & 4 & -0.48 & 0.24 & 7 \\
Be 70 & 12.5 & 4 & -0.32 & - & 1 \\
Be 73 & 17.4 & 1.5 & -0.22 & 0.05 & 5 \\
Be 75 & 15.5 & 4 & -0.22 & 0.2 & 5 \\
Cr 261 & 6.96 & 6 & -0.03 & 0.03 & 4 \\
IC 4651 & 7.7 & 1.7 & 0.1 & 0.03 & 2 \\
King 5 & 10 & 1 & -0.38 & 0.17 & 1 \\
King 11 & 10.1 & 4.13\footnotemark[$\dagger$] & -0.23 & 0.15 & 1\\
Melotte 66 & 10.2 & 4 & -0.38 & 0.15 & 2 \\
Melotte 71 & 10 & 1 & -0.32 & 0.16 & 2 \\
NGC 1193 & 12.2 & 7.9 & -0.29 & 0.17 & 1 \\
NGC 188 & 9.6 & 4.28 & -0.02 & 0.09 & 1 \\
NGC 2112 & 9.2 & 2 & -0.1 & 0.2 & 2 \\
NGC 2141 & 11.8 & 2.5 & -0.18 & 0.15 & 3 \\
NGC 2243 & 10.2 & 4 & -0.48 & 0.15 & 4 \\
NGC 2360 & 9.3 & 1 & 0.07 & 0.17 & 2 \\
NGC 2420 & 10.6 & 2 & -0.47 & 0.07 & 2 \\
NGC 2506 & 10.38 & 1.7 & -0.2 & 0.03 & 4 \\
NGC 2660 & 8.68 & 1 & 0.04 & 0.04 & 6 \\
NGC 2682 & 9.1 & 2.56 & 0 & 0.05 & 1 \\
NGC 3680 & 8.2 & 1.19 & -0.09 & 0.05 & 1 \\
NGC 3960 & 7.8 & 1 & 0.02 & 0.04 & 6 \\
NGC 5822 & 7.8 & 1.2 & 0.1 & 0.1 & 2 \\
NGC 6208 & 7.6 & 1.17 & -0.03 & 0.06 & 1 \\
NGC 6253 & 6.6 & 3 & 0.46 & 0.11 & 8 \\
NGC 6791 & 8.1 & 9 & 0.47 & 0.12 & 9 \\
NGC 6819 & 7.71 & 3 & 0.09 & 0.03 & 4 \\
NGC 6939 & 8.7 & 2.21 & 0.02 & 0.11 & 1 \\
NGC 7142 & 9 & 1.88 & 0.04 & 0.01 & 1 \\
NGC 752 & 8.8 & 1.12 & -0.08 & 0.07 & 1 \\
NGC 7789 & 9.4 & 1.5 & -0.04 & 0.09 & 2 \\
Rup 46 & 8.9 & 3.98 & -0.04 & 0.18 & 1 \\
To 2 & 15.1 & 2.2 & -0.28 & 0.05 & 10 \\
Saurer 1 & 19.3 & 4 & -0.38 & 0.14 & 2 \\
M 67 & 8.6 & 4 & 0.02 & 0.14 & 3 \\
\hline
\multicolumn{6}{@{}l@{}}{\hbox to 0pt{\parbox{80mm}{\scriptsize \footnotemark[$*$] The distance of Be 18 is taken from \citep{Friel_95}. See also \citet{Carraro_99}. \footnotemark[$\dagger$] The age of King 11 is taken from \citet{Tosi_07}. References.$-$(1) \citet{Chen_03}; (2) \citet{Friel_06}; (3) \citet{Yong_05}; (4) \citet{Bragaglia_06}; (5) \citet{Carraro_07}; (6) \citet{Sestito_06}; (7) \citet{Villanova_05}; (8) \citet{Carretta_07}; (9) \citet{Gratton_06}; (10) \citet{Frinchaboy_08}. }\hss}} 
\end{tabular}
}
\end{center}
\end{table}


\end{document}